\documentclass{PoS}
\usepackage{amsmath,amsfonts}
\usepackage{graphics}
\usepackage{epsfig}
\usepackage{feynmp,psfrag}
\PoS{PoS(LAT2005)011}

\title{High Precision Fundamental Constants using Lattice Perturbation Theory}

\ShortTitle{High Precision Fundamental Constants using Lattice Perturbation Theory}

\author{\speaker{Quentin Mason}\thanks{For the HPQCD Collaboration.}\\
         University of Cambridge\\
        E-mail: \email{Q.J.Mason@damtp.cam.ac.uk}}

\author{Howard Trottier\\
        Simon Fraser University}%
\author{Ron Horgan\\
         University of Cambridge}%\\
\author{HPQCD Collaboration}
\abstract{The HPQCD collaboration has a program for determining the fundamental constants of the Standard Model Lagrangian from lattice QCD.  The most accurate method 
of doing this uses the $n_f=2+1$ improved staggered MILC ensembles with chiral fitting and multi-loop perturbative renormalisation to connect to the continuum \msbar\ scheme.
This program has already been very successful with the recent strong coupling constant determination at three-loops from 28 observables at three lattice spacings, and the one-loop light quark mass calculation last year.  
Here a preliminary result is presented for the first-ever lattice determination
of the two-loop multiplicative quark mass renormalisation.  
The perturbative calculation involved was automated in the generation of the Feynman rules, and the generation and coding of all of the roughly 30 Feynman diagrams.  
The full formal framework for lattice quark mass renormalisation is given, including the cancellation of infrared divergences in intermediate diagrams.
The result was checked by evaluation in three separate gauges and by 
two authors independently, showing the incredible flexibility and power 
of this perturbative methodology. Our preliminary result for the two-loop
perturbative matching factor, and of systematic errors associated with
higher-orders, gives \msbar\ masses at a 2~GeV scale of 
$m_s = 87(0)(4)(4)(0)$~MeV, and $\frac12(m_u+m_d) = 3.3(0)(2)(2)(0)$~MeV,
where the respective uncertainties are from lattice statistical, lattice systematic, 
perturbative, and electromagnetic and isospin effects. The perturbative errors
are a factor of two smaller than in our previous study, and
we anticipate reducing this somewhat further from additional analysis 
of the systematics.}

\FullConference{XXIIIrd International Symposium on Lattice Field Theory\\
                 25-30 July 2005\\
                 Trinity College, Dublin, Ireland}
\newcommand{\msbar}{\text{$\overline{\text{MS}}$}}
\newcommand{\naive}{\text{na\"\i ve}}

\DeclareMathOperator{\Tr}{Tr}
\DeclareMathOperator{\SIGMA}{\Sigma}

\def\et{{\it et al.}}
\def\schpt{S\raise0.4ex\hbox{$\chi$}PT}

\begin{document}
\section{Introduction}
The strong sector of the Standard Model Lagrangian contains a number of inputs that are \emph{a priori} unknown and must be determined from experiment.  For these fundamental constants; the quark masses and the strong coupling constant, our knowledge is currently rather imprecise.  Their determination is complicated by confinement in QCD, so the quarks and gluons cannot be observed as isolated particles.  We can only determine these fundamental parameters by solving QCD for observable quantities 
such as hadron masses, as functions of the quark masses and the coupling 
(or alternatively, the lattice spacing).  The High Precision QCD (HPQCD) collaboration has a program for calculating the values of these parameters using the numerical techniques of lattice QCD simulations combined with multi-loop perturbative renormalisation. Previously the masses have been determined at the one-loop level~\cite{ms,matt,Gray:2005ur}, and the strong coupling constant was determined from one observable in a two-loop calculation~\cite{Davies:2003ik}.  Recently the 
determination of the strong coupling was improved with three-loop perturbative
matching and with input from 28 lattice observables in simulations at three different lattice spacings, resulting in an accuracy of just over 
1\%~\cite{Mason:2005zx,AlphaPT}. This writeup covers progress on the 
determination of the light quark masses, where we push the
perturbative matching calculation to two loops, that is, 
next-to-next-to-leading order.

Precise knowledge of quark masses constrains Beyond 
the Standard Model scenarios as well as providing 
input for phenomenological calculations of Standard 
Model physics.
The strange quark mass, in particular, is needed for various phenomenological 
studies, including the important CP-violating quantity 
$\epsilon^\prime/\epsilon$~\cite{Buras:1996dq}, where 
its uncertainty severely limits the theoretical precision.

Previously, shortcomings in the formulation of 
QCD on the lattice and limitations in computing power have 
meant that lattice calculations were forced to work 
with an unrealistic QCD vacuum that either 
ignored dynamical (sea) quarks or included only 
$u$ and $d$ quarks with masses much heavier than in nature. 
This condemned determinations of most phenomenologically-important
quantities, including the quark masses, 
to rather large systematic errors (10--20\%) arising from the 
inconsistency of comparing such unrealistic theories with 
the necessary experimental input. The determination presented here 
uses simulations with the improved staggered quark formalism that 
generates a much more realistic QCD vacuum with two light dynamical 
quarks and one strange dynamical quark. Staggered quarks are fast to 
simulate. They keep a remnant of chiral symmetry on 
the lattice, which prevents the occurrence of exceptional configurations,
and which allows simulations at much smaller quark masses.
The bare quark masses in the simulations were fixed using chiral 
perturbation theory to reliably extrapolate lattice results to the physical 
point~\cite{ms}. Working in the region of dynamical $u/d$ 
quark masses below $m_s/2$ and down to $m_s/8$ gives control 
of chiral extrapolations and avoids the large systematic errors from dynamical 
quark mass and unquenching effects that afflicted previous studies
using other lattice discretisations. 

The dominant systematic error 
in the determination of the \msbar\ masses in~\cite{ms} came from 
unknown second- and higher-orders in the perturbative matching.
Some progress was reported on the chiral fits at the 
lattice meeting~\cite{claude_poster}, however that analysis is not used here;
we continue to employ the bare quark masses given in~\cite{ms}.
Significant progress on the reduction of the systematic errors is
reported here, due to our computation of the second-order perturbative 
matching coefficient, the first determination at this order of a 
``kinetic'' quark mass in any lattice theory (the zero-point 
additive renormalisation for Wilson fermions was previously 
determined at two-loops in~\cite{Panagop}, and for static quarks 
in~\cite{Martinelli:1998vt}).

The staggered quark formalism does present several challenges, 
which have been tamed with an aggressive program of
perturbative improvement. With the \naive\ staggered action
large discretisation errors appear, although they 
are formally only $\mathcal{O}(a^2)$ or higher ($a$ is the lattice spacing). 
In the case of the unimproved staggered action
the renormalisation of lattice operators to match to continuum 
quantities were also frequently large and poorly convergent in perturbation 
theory. This was true, for example, for the mass 
renormalisation that is needed here. It turns out 
that both problems have the same source, a particular form 
of discretisation error in the action, called ``taste violation,'' and 
both are ameliorated by use of 
the improved staggered formalism~\cite{Lepage:1998vj}. 
The perturbation theory then shows small 
renormalisations~\cite{Hein:2001kw,Lee:2002ui,Trottier:2003bw,Mason:2002mm} 
and discretisation errors are much reduced
~\cite{Blum:1997uf,Bernard:1998mz,Orginos:1998ue}.
Empirically, taste violation remains
the most important discretisation error in the improved theory, despite being 
subleading to ``generic'' discretisation errors. The Goldstone meson masses
we will discuss here are affected by this 
at one-loop in the chiral expansion. 
Staggered chiral perturbation theory
(\schpt)~\cite{Lee:1999zx,Bernard:2001yj,Aubin:2003mg,Aubin:2003ne}
allows us to control these effects and reduce discretisation errors significantly.

A potentially more fundamental concern about staggered fermions relates 
to the need to take the fourth root of the quark determinant, in order to 
convert the four-fold duplication of ``tastes'' into one quark flavor. 
One might imagine that the fourth root introduces nonlocalities which prevent 
decoupling of the ultraviolet modes of the theory in the continuum limit.
However, evidence is amassing that demonstrates that the properties of the staggered theory, with the
fourth root, are equivalent to a one-flavor theory, up to the expected
discretisation errors.  These are due to short-distance taste-changing 
interactions, which are mediated by high-momentum gluons~\cite{Lepage:1998vj} 
(the locality of the free-field staggered theory is trivial, and is made
manifest in the ``naive'' basis used in~\cite{Lepage:1998vj}). One should
not be surprised that nonlocalities do not arise, precisely because the
staggered quark matrix is diagonal in the taste basis, up to those small,
short-distance (and calculable) corrections. 
%It is also clear that the 
%smallness of the off-diagonal components of the staggered matrix in the 
%taste basis, and hence the locality of the fourth-root of its determinant, 
%hold non-perturbatively, at least for gauge fields that are sufficiently 
%smooth at the scale of the cutoff. In this connection, one may note that an 
%analytical non-perturbative proof of the locality of the overlap operator 
%is also only available in the limit of sufficiently smooth gauge fields 
%\cite{Hernandez:1998et}. 
%With respect to realistic gauge-field backgrounds,
It has been demonstrated that perturbative improvement of staggered actions 
correlates exceedingly well with non-perturbatively measured properties 
of the staggered fermion matrix, providing clear support for the correctness
of the fourth-root procedure. This includes the measured pattern of
low-lying eigenvalues of the staggered matrix 
\cite{Follana:2004sz,Wong:2004nk,Durr:2004as}, 
and the measured pattern of taste-violating mass differences in the 
non-chiral pions~\cite{Follana:2004mg}.

The rest of this paper is organised as follows.
The methodology of the calculation is discussed in the following section,
including a general discussion of how to obtain the matching factor
which connects the bare quark mass $m_0(a)$ to the
\msbar\ mass $m^\msbar(\mu)$, using the pole mass as an intermediate
quantity. We also derive the two-loop anomalous dimension for the bare
quark mass in the lattice-regularised theory, and give complete 
results for the two-loop matching coefficients.
Section~\ref{S:Results} gives preliminary results for the light quark masses, 
with a preliminary analysis of the systematic uncertainties,
including a technique to make a rough estimate of the third-order 
perturbative correction. Section~\ref{S:Discussion} compares our
results with other recent determinations of the strange-quark mass.
An appendix provides some additional information concerning the
evaluation of the multi-loop diagrams, including explicit expressions 
for the two-loop quark mass renormalisation in terms of the 1PI
self-energy, a discussion of the techniques used for generating
and evaluating the necessary multi-loop integrands, and some
detailed numerical results which provide an indication of the
many consistency checks that we have applied to our results.

\section{Structure of Calculation}
The quark masses are not physically measurable, and as such are only well-defined in certain renormalisation schemes, such as the \msbar\ mass $m^\msbar(\mu)$,
evaluated at some convenient scale $\mu$. The light quark \msbar\ masses are
determined here by multiplying the chirally-extrapolated bare masses in
lattice units, $a m_0$, by the inverse lattice spacing $a$, and by the 
appropriate perturbative matching factor $Z_m(\mu a,m_0 a)$, which we 
compute to two-loop order:
\begin{equation}
m^\msbar(\mu) = \frac{(m_0\, a)}{a}\; Z_m(\mu a,m_0 a),
\label{e:Zm}
\end{equation}
where the bare mass $m_0(a)$ is cutoff dependent.
The bare masses used here were determined in an extensive 
chiral perturbation theory analysis of the MILC Asqtad data that was 
discussed in~\cite{ms,MILC_SPECTRUM,MILC_chiralPT,Gottlieb:2003bt}. 
These bare masses were previously
used with the one-loop perturbation theory result for $Z_m$ to 
extract the \msbar\ masses in~\cite{ms};  
the final result for the strange quark mass reported there 
was 76(0)(3)(7)(0)\,MeV where the respective errors are 
from: statistics; simulations systematics of which the most important are chiral fitting and lattice spacing; an estimate of the unknown two-loop perturbative errors and an 
estimate of the uncertainty due to electromagnetic and isospin contributions to the pion and kaon.  That the error
coming from the lattice spacing $a$ is so small is a distinguishing feature of this calculation.  
The lattice spacing~$a$ is one of the five simulation parameters, and an important one because it sets the simulation's mass scale. In our earlier light quark masses and $\alpha_s$~analyses,   we set the lattice spacing by comparing a simulated $\Upsilon$ mass splitting (\emph{e.g.}, $\Upsilon^\prime-\Upsilon$) with experiment. Here we continue this practice, although the lattice spacings derived from our $\Upsilon$~splitting have been shown to agree with spacings derived from a wide variety of other physical quantities: ten in all, including the pion and kaon leptonic decay constants, the $B_s$~mass, and the $\Omega$~baryon mass\,\cite{Davies:2003ik,omega}. All of these different quantities agree on the lattice spacing to within~1.5--3\%. 

The largest error in our previous determination of the quark masses
\cite{ms} was from the perturbative matching.  
Here that is addressed by the calculation of $Z_m$ at two loops.
We do this in two stages, using the pole mass $M$ as a matching quantity
to connect the lattice- and \msbar-regularisation schemes. We also
use our previous determination of the relation between the lattice bare 
coupling and the renormalised coupling $\alpha_V(q^*)$, defined by the
static potential, to reorganise both sides of the matching equation
into series in terms of $\alpha_V(q^*)$ at an appropriately determined scale.

We begin by recalling the relation between the \msbar\ mass
and the pole mass $M$, which is known through three loops 
\cite{Tarrach:1980up,Gray:1990yh,Broadhurst:1991fy,%
Chetyrkin:1999qi,Melnikov:2000qh}.
We require it to second order, a result that was first obtained in 
\cite{Gray:1990yh} (expressions for the relation at arbitrary $\mu$
are conveniently given in~\cite{Chetyrkin:1999qi})
\def\ellu{\ell_{\mu M}}
\def\sfrac#1#2{{\textstyle{\frac{#1}{#2}}}}
\begin{align}
   m^\msbar(\mu) = M \left[ 1 +    
  z_1\left(\frac{\mu}{M}\right)\frac{\alpha_\msbar(\mu)}{\pi} 
+ z_2\left(\frac{\mu}{M}\right)\frac{\alpha^2_\msbar(\mu)}{\pi^2} + \ldots \right],
\label{e:msbarMpole}
\end{align}
where the one- and two-loop coefficient functions
$z_1(\mu/M)$ and $z_2(\mu/M)$ are reduced to a set of terms
with different colour structures [in the following 
$C_F = (N_c^2 - 1)/(2 N_c)$, $C_A = N_c$, and $T=1/2$]
\begin{align}
   z_1 &= C_F z_F \\
   z_2 &= C_F^2 z_{FF} + C_F C_A z_{FA} + C_F T n_\ell z_{FL} + C_F T z_{FH} ,
\label{e:z12}
\end{align}
and where the contribution $z_{FH}$ from an internal quark loop with the
same flavor as the valence quark is split off from the contribution $z_{FL}$
of $n_\ell$ internal quark loops with different flavor (these are taken here to
be degenerate in mass, though this is easily generalised). The total number of
flavors is $n_f = n_\ell + 1$. The individual functions are given by
\begin{align}
z_F&=-1-\sfrac34\ellu ,\\
z_{FF}&=\sfrac{7}{128}-\sfrac{15}{8}\zeta_2-\sfrac34\zeta_3+3\zeta_2\log 2
+\sfrac{21}{32}\ellu+\sfrac{9}{32}\ellu^{\,2} ,\\ 
z_{FA}&=-\sfrac{1111}{384}+\sfrac12\zeta_2+\sfrac38\zeta_3-\sfrac32\zeta_2\log 2
-\sfrac{185}{96}\ellu-\sfrac{11}{32}\ellu^{\,2} ,\\ 
z_{FL}&=\sfrac{71}{96}+\sfrac12\zeta_2-2\Delta(r_{\rm sea})
+\sfrac{13}{24}\ellu+\sfrac18\ellu^{\,2}, \label{e:zFL} \\
z_{FH}&=\sfrac{71}{96}+\sfrac12\zeta_2-2\Delta(1)
+\sfrac{13}{24}\ellu+\sfrac18\ellu^{\,2}, \label{e:zFH}
\end{align}
where $\ellu \equiv \log(\mu^2/M^2)$, and where the function $\Delta(r)$ 
gives the dependence of the renormalisation factors $z_{FL}$ 
and $z_{FH}$ on the quark mass in an internal fermion loop (sea and valence,
respectively), with $r_{\rm sea} = m_{\rm sea}/m_{\rm valence}$. 
An exact integral expression for $\Delta(r)$ can be found in 
\cite{Gray:1990yh} and it is plotted in the appendix. %Particular limits are~\cite{Gray:1990yh,Broadhurst:1991fy}
%\begin{align}
%&\Delta(r \ll 1) = \sfrac34 \zeta_2 r + O(r^2) , \label{e:Deltall} \\
%&\Delta(1) = \frac{\pi^2-3}{8} ,\\ 
%&\Delta(r\gg1) = \sfrac14 \log^2\,r + \sfrac{13}{24}\log\,r + \sfrac14 \zeta_2 + \sfrac{151}{288} + O(r^{-2} \log\,r) . \label{e:Deltagg}
%\end{align}

Next we outline the computation of the relation between the pole mass $M$ 
and the bare quark mass $m_0(a)$ in the lattice-regularised theory, which 
has the form
\begin{align}
M = m_0\left[1 + \alpha_L \left(A_{11}\log m_0 a+A_{10}\right) 
+ \alpha_L^2 \left( A_{22}\log^2m_0a + A_{21}\log m_0 a+A_{20} \right)+\ldots\right] .
\label{e:LatticeMpole}
\end{align}
The coefficients of the logarithms are determined by the two-loop
anomalous dimension in $m_0(a)$, which in turn can be determined 
from the known anomalous dimension of the $\msbar$ mass, as described
below. We have previously computed the one-loop matching factor, with
the result~\cite{ms,Hein:2001kw} $A_{10} \approx 0.5432$
(neglecting corrections of $O((a m_0)^2)$). What is new in this paper is the determination 
of the two-loop term $A_{20}$.

As with the two-loop continuum matching factor in~\eqref{e:msbarMpole}, $A_{20}$ depends on the number of quark 
flavors in the sea, and on the ratios of the sea quark masses
$m_{\rm sea}$ to the valence quark mass $m_{\rm valence}$. However, 
as we demonstrate explicitly in the appendix,
the mass dependence in $A_{20}$ cancels precisely in the matching to 
the continuum relation~\eqref{e:msbarMpole} for $m_{\text{sea}}/m_{\text{valence}}$ in the range necessary.
%, in the limit where the sea 
%and valence quark masses are all much less than the lattice ultraviolet 
%cutoff $\pi/a$, and the \msbar\ scale $\mu$. This cancellation holds
%for an arbitrary ratio $m_{\rm sea}/m_{\rm valence}$ [compare with the 
%mass dependent corrections in the continuum quantities $z_{FL}$ and 
%$z_{FH}$ in Eqs. (\ref{e:zFL}) and (\ref{e:zFH})].
This follows from the fact that, in the limit where the energy scales
are large, the net renormalisation factor connecting $m_0(a)$ to 
$m^\msbar(\mu)$ probes internal loops at scales large compared to 
the internal quark masses; hence the 
dependence on quark masses of the intermediate renormalisation 
factors connecting to the pole mass $M$ are infrared effects 
that are identical in the lattice- and \msbar-regularised theories.

Before we give our results for $A_{20}$, and the final matching
factor $Z_m$ in~\eqref{e:Zm}, we consider how to determine the 
coefficients of the logarithmic terms in~\eqref{e:LatticeMpole}.
The pole mass is an RD invariant (equivalently does not depend on $a$) 
so taking logs and differentiating with respect to $\log a$, and 
neglecting any scale dependence in the coefficients $A_{nm}$
(which would arise as discretisation errors that go as powers of 
$m_0 a \lesssim 0.05$, which are negligible for our purposes), one has
\begin{align}
0=\frac{d \log m_0}{d \log a} + \frac{d}{d\log a}\Biggl\{\Bigl[\cdots\Bigr]-\frac12\Bigl[\cdots\Bigr]^2\Biggr\},
\label{e:RG}
\end{align}
where the square bracket is from \eqref{e:LatticeMpole}, and the first
term above can be identified with the anomalous dimension equation:
\begin{align}\label{e:latanondim}
\frac{d \log m_0}{d \log a} \equiv \gamma_0\,\alpha_L(a)+\gamma_1^L\,\alpha_L^2(a) + \mathcal{O}(\alpha_L^3)
\end{align}
The $L$'s indicate the lattice-regularisation scheme.  The sign comes from differentiating with respect to the logarithm
of the lattice scale $a^{-1}$ (momentum units), instead of $\mu$.
Only the leading term in the anomalous dimension is scheme independent,
$\gamma_0 = 3C_F/(2\pi)$. Comparison of the two preceding equations
leads to the identification
\begin{align}
A_{11}=-\gamma_0, \qquad
A_{22}=\frac12A_{11}^2-A_{11}\frac{\beta_0}{4\pi}, \qquad
A_{21}=-\gamma_1^L+A_{11}^2-A_{10}\left(\frac{\beta_0}{2\pi}-A_{11}\right),
\end{align}
where the $\beta$-function arises from an implicit derivative of
the bare lattice coupling in~\eqref{e:RG}
($\beta_0 = 11-\frac23n_f$). It remains to determine $\gamma_1^L$.
This can be found from the known \msbar\ anomalous dimension,
\begin{align}
-\frac{d \log m_\msbar}{d \log \mu} &\equiv \gamma_0\alpha_\msbar(\mu)+\gamma_1^\msbar\alpha_\msbar^2(\mu)+\mathcal{O}(\alpha^3),
\qquad \gamma_1^\msbar=\frac{101}{12\pi^2}-\frac{10}{36\pi^2}n_f,
\end{align}
making the substitution $\mu=a^{-1}$, and using the relations
$m_\msbar(a^{-1})=m_0(1+C_m\alpha)$, and $\alpha_\msbar(a^{-1})=\alpha_L(1+C_\alpha\alpha_L)$. By comparison to~\eqref{e:latanondim}
we obtain
\begin{align}
\gamma_1^L&=\gamma_1^\msbar+C_\alpha\gamma_0-C_m\frac{\beta_0}{2\pi} .
\end{align}
The one-loop renormalisation constants $C_m = Z_m^{(2)}(\mu=1/a) = 0.1188$ and 
$C_\alpha = 4.753-0.3316n_f$ for the improved gluon and staggered quark actions
have been previously calculated~\cite{ms,Hein:2001kw,Mason:2005zx,AlphaPT}.

\begin{figure}[t]
\begin{center}
\includegraphics[width=5in]{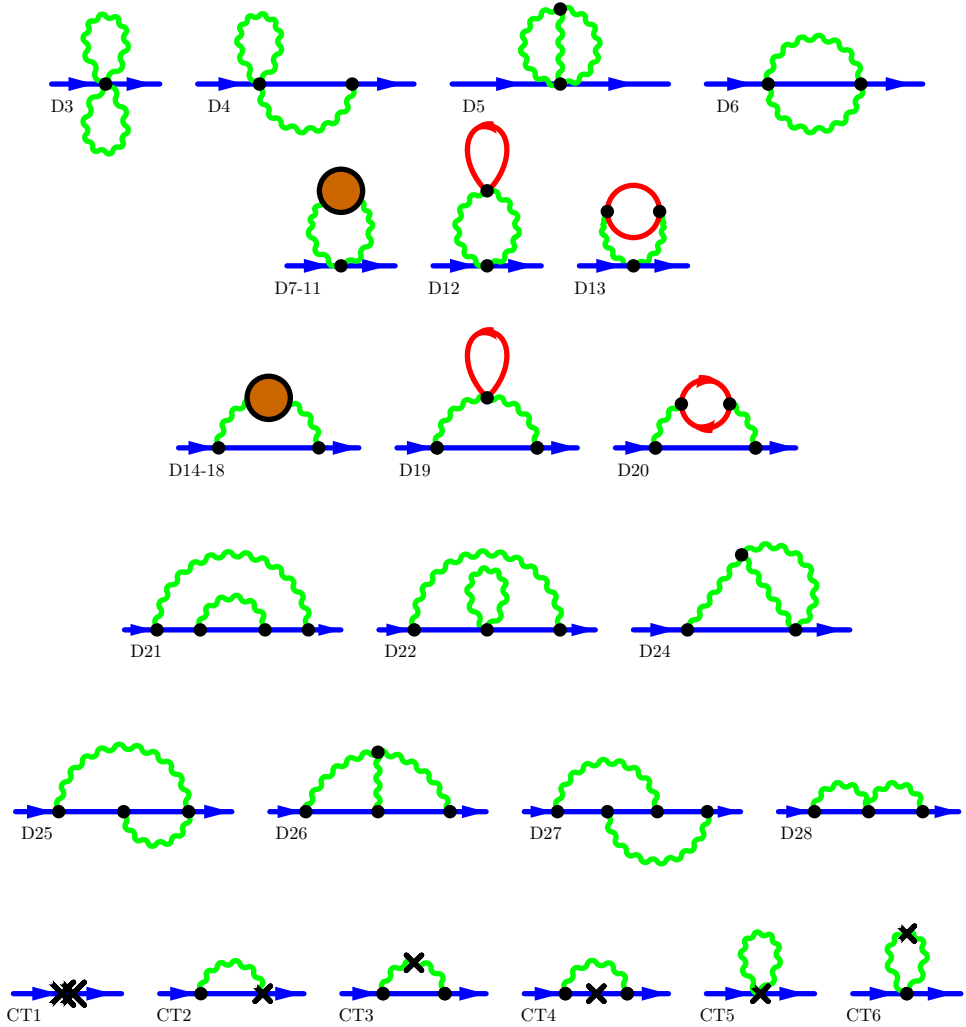} %4.5in
\caption{\label{f:QMO2diags}The two-loop diagrams that contribute to 
the two-loop mass renormalisation.  The numbering is consistent with \protect\cite{Panagop}.  Blue lines are valence quarks (on-shell externally), green are gluons and red are sea quarks.  The large brown vertices are stand-ins for the five one-loop gluon propagator diagrams (gluon and ghost bubbles and tadpoles and the measure term).  We evaluate all five simultaneously with an internal subtraction. The crosses represent interaction vertices generated
by perturbative expansion of tadpole and other renormalisation factors in the
gluon and quark actions.}
\end{center}
\end{figure}

Knowing the logarithmic terms due to the lattice anomalous dimension 
provides a useful cross-check on our evaluation of the two-loop 
renormalisation factor~\eqref{e:LatticeMpole}. We compute the
two-loop lattice diagrams as a function of $m_0 a$, and subtract 
the known logarithms, in order to isolate the remaining term $A_{20}$, 
which must be finite as $m_0 a \to 0$.
Additional checks are provided for diagrams which have a leading
$\log^2(am_0)$ term, which arises from the infrared limit of both 
the outer and inner loop integrals, and which is therefore an infrared 
quantity; the coefficients
of the double logarithms in the individual diagrams are available in 
Feynman gauge from the original \msbar\ calculation of 
Tarrach~\cite{Tarrach:1980up}, and we have verified
that these are reproduced in our lattice calculation.   

The necessary two-loop diagrams are shown in figure~\ref{f:QMO2diags}.
Further details on our evaluation of these diagram are given in the appendix.
We can easily evaluate them for various gluon and quark actions, using
our automated methods for generating the lattice Feynman rules. In
this paper we give results for the Symanzik improved gluon with improved 
staggered quarks (Asqtad) and SU(3). Our result for the matching term $A_{20}$ is,
in the limit of vanishing sea quark mass,
\begin{align}
   A_{20} = 6.09 - 0.15 n_\ell - 0.03 ,
%   A_{20} = 6.092(4) - 0.1484(3) n_\ell - 0.0328(2) ,
\label{e:A20}
\end{align}
where the last term corresponds to an internal quark loop containing one (massive) 
valence quark flavor. The uncertainties arise from a numerical evaluation
of the two-loop integrals.

When the lattice renormalisation factor~\eqref{e:LatticeMpole} connecting 
the bare mass to the pole mass is combined with the equivalent continuum
expression~\eqref{e:msbarMpole} for the connection to the \msbar\ mass, 
the logs of $m_0$ drop out, as expected, since these are infrared effects 
that are identical in the intermediate lattice and continuum matchings.
We also reorganise the couplings to the $\alpha_V$
scheme at some scale $q^*$, where $q^* a$ is a function of $\mu a$
that is determined according to the BLM scheme~\cite{BLM}.
This leaves an expression with logarithms only of $\mu a$ and $q^* a$.
The final expression for the perturbative matching factor is then:
\begin{align}
Z_m(\mu a,m_0 a) = 1 + Z_m^{(2)}(\mu a)\alpha_V\left(q^*(\mu a)\right)
+Z^{(4)}_m(\mu a)\alpha_V^2+\mathcal{O}((m_0 a)^4,(m_0 a)^2\alpha^2,\alpha^3),
\label{e:Zmfinal}
\end{align}
where $Z_m^{(2)}(\mu a) = 0.1188(1)-2/\pi\log(\mu a)+\mathcal{O}(m_0 a)^2$ was derived 
previously,~\cite{ms}, and the new information presented here is the 
expression for $Z_m^{(4)}$
\begin{align}
Z_m^{(4)}(\mu a) = Z_2 \log^2(\mu a) + Z_1 \log(\mu a) + Z_0,
\end{align}
where the latter coefficients depend on logarithms of $q^* a$
\begin{align}
Z_2 &= 0.76-0.034 n_f\\
Z_1 &= -0.40-0.028 n_f +(-1.12+0.0675 n_f)\ln q^*a\\
Z_0 &= 2.09-0.014n_f+(0.21-0.013n_f)\ln q^*a.
%Z_2 &=  0.7599 - 0.0338 n_f ,\\
%Z_1 &= -0.4049 - 0.0281 n_f + (-1.1145 + 0.0675 n_f)\log(q^*a) ,\\
%Z_0 &= 2.086(4) - 0.0144(3) n_f + (0.2080-0.0126n_f)\log(q^*a).
\end{align}
We have computed the optimal choice for $q^*a$ in the BLM scheme, 
as a function of $\mu a$,
using an exact evaluation of the average momentum scales in the continuum
self-energy diagrams, according to~\cite{BLM} (this is an improvement over the inexact calculation in~\cite{ms}), and a numerical
evaluation of the average scales in the lattice self-energy (unchanged). 
A typical value for the matching scale is $q^*a = 1.88$ at $\mu a=1$.

\section{Results}\label{S:Results}
The bare lattice masses for the strange and up/down quarks, on the
MILC ``coarse'' and ``fine'' lattices, are given in~\cite{ms}.
For the strange quark, these are $a m_{0s} = 0.0390(1)(20) / u_{0c}$,
and $a m_{0s} = 0.0272(1)(12) / u_{0f}$, on the coarse and fine
lattices respectively, where $u_{0c}=0.85488$ and $u_{0f}=0.86774$ are tadpole
normalisation factors. The uncertainties are lattice statistical
and systematic errors, respectively, the latter
due mainly to chiral extrapolation/interpolation. The lattice spacings 
can be found in~\cite{Mason:2005zx},
$a^{-1}_{\rm coarse}=1.596(30)$~GeV, and $a^{-1}_{\rm fine}=2.258(32)$~GeV.

Following conventional practice, we quote the light quark \msbar\ masses
at the scale $\mu=2$~GeV, taking three active flavors of quarks ($n_f=3$).
%The BLM scales on the two MILC lattices are then
%$a q^*_{\rm coarse} = 2.144$ and $a q^*_{\rm fine} = 1.752$ (these
%values are more accurate than the slightly larger values given in
%\cite{ms}, where an approximate determination was made of the average
%loop momentum circulating in the continuum self-energy).
This results in two-loop coefficients in~\eqref{e:Zmfinal}
of $Z_m^{(4)}\vert_{\rm coarse} = 1.939(4)$, 
and $Z_m^{(4)}\vert_{\rm fine} = 2.270(4)$.
Finally, we require the couplings at the relevant scales,
and for this purpose we use the recently determined value 
$\alpha_V^{(n_f=3)}(7.5~\mbox{GeV}) = 0.2082(40)$~\cite{Mason:2005zx}.
We find $\alpha_V(q^*_{\rm coarse}) = 0.2925(92)$ and
$\alpha_V(q^*_{\rm fine}) = 0.2713(76)$.

Putting all of this together, we obtain the following preliminary values for the 
\msbar\ strange-quark mass, using the coarse and fine lattices as input
\begin{align}
   m^\msbar_s(2~\mbox{GeV}) = 83(5)~\mbox{MeV\ [coarse]}, \quad
   m^\msbar_s(2~\mbox{GeV}) = 85(4)~\mbox{MeV\ [fine]} ,
\label{e:msvalues}
\end{align}
where the errors here are just from the simulation systematics.

Following~\cite{ms}, we consider continuum extrapolations of the
central values in \eqref{e:msvalues} based on the form of the 
expected leading discretisation errors, which are of $O(\alpha_V a^2)$,
and compare with a extrapolation in $\alpha_V^2 a^2$. These two forms
yield almost identical extrapolations, to a central value of 87~MeV, with an error
that is folded into the quoted lattice systematic errors above.
We are currently considering refinements to our estimates of 
the continuum extrapolation and the incorporation of third order perturbative errors. 
For this preliminary result we quote
as our continuum extrapolation
\begin{align}
  m^\msbar_s(2~\mbox{GeV}) = 87(0)(4)(4)(0)~\mbox{MeV}
\end{align}
where, following~\cite{ms}, the respective errors are statistical, lattice
systematic, perturbative, and electromagnetic. 
We have assigned a relative error from the estimated third-order 
perturbative matching of roughly $\pm 2 \alpha^3_V(q^*) \approx 5\%$, 
down from the 10\% perturbation theory error in the one-loop 
determination in~\cite{ms}. We anticipate to reduce somewhat 
the perturbation theory error with a more careful 
study of the systematics of the $\mu$ dependence.

Our result for the ratio of strange to up/down quark masses is
unchanged at 27.4(1)(4)(0)(1) from~\cite{ms}, since the renormalisation factor is
mass-independent, as we have verified above explicitly through 
two-loops.
%, hence we quote
%\begin{align}
%  \frac{m_s}{\hat m} = 27.4(1)(4)(0)(1) ,
%\end{align}
%where $\hat m \equiv \frac12(m_u + m_d)$. Alternatively, we have
Using the value $m_u/m_d=0.43(0)(8)(1)(0)$ from~\cite{MILC_chiralPT} this gives:
\begin{align}
  m^\msbar_u(2~\mbox{GeV}) = 1.9(0)(1)(1)(2)~\mbox{MeV}\notag\\
  m^\msbar_d(2~\mbox{GeV}) = 4.4(0)(2)(2)(2)~\mbox{MeV}\notag
\end{align}

\section{Discussion and Conclusions}\label{S:Discussion}
Perturbation theory has once again shown itself to be an essential tool in high precision phenomenological calculations from the lattice. The two-loop lattice diagrams were conquered with a combination of algebraic and numerical
techniques in this first-ever determination of the multiplicative two-loop 
pole mass on the lattice.  When combined with the known continuum matching 
from the pole mass to the
\msbar\ mass, a very accurate determination of the light quark masses 
was possible.  The results presented here have a number of 
distinguishing features:  two-loop perturbation theory, $n_f=2+1$ simulations with two degenerate light quarks and a heavier strange quark, 
very small light quark masses from $m_s/8$ to $m_s/2$ which enabled a partially quenched chiral fit with many terms to thousands of configurations, and
extremely accurate determinations of the lattice spacings, which are
equal within the errors when set from any of 10 different 
quantities, ranging from the very lightest states all 
the way up to heavy mesons and baryons~\cite{Davies:2003ik,omega}.

Most notable amongst our results is our new value for strange quark mass
$m^\msbar_s(2~\mbox{GeV}) = 87(0)(4)(4)(0)~\mbox{MeV}$, where
the respective errors are lattice statistical, lattice systematic 
(mostly due to the chiral extrapolation/interpolation), perturbative, 
and due to electromagnetic effects. The two-loop matching has increased 
the central value with respect to the previous determination in~\cite{ms} 
by about two standard deviations, based on the previous estimate of the perturbation theory uncertainty (which was $\pm 7$~MeV). 
We believe the present estimate of the perturbative uncertainty of 
$\pm 4$~MeV can be reduced somewhat by estimating the third-order 
perturbative correction, along the lines that we have already
explored, in a preliminary way, here. 

The strange quark mass determination has historically generated some
controversy, with somewhat different values being obtained from different
approaches. An obvious advantage of our result is that it has been
obtained with the correct description of the sea, that is, with
$n_f=2+1$ flavors of dynamical quarks. There is only one other result with the 
correct number of flavors in the sea, which is due to the CP-PACS and
JLQCD collaborations, who reported a value at this conference
of $m^\msbar_s(2~\mbox{GeV}) = 87(6)$~MeV~\cite{CPPACS-JLQCD},
although the error was not very well quantified, and does not appear to 
include an estimate of corrections due to two-loop and higher-orders in the perturbative matching.

It appears the the strange-quark mass extracted from simulations with
only $n_f=2$ flavors of sea quarks are systematically higher than the
estimates with the correct $n_f$ (noting that the two-flavor 
determinations were also done with nonperturbative definitions 
of the quark mass),
although the other systematic errors are too large to allow for a 
definitive assessment. The two-flavor determination from the 
QCDSF-UKQCD collaboration, 
is $m^\msbar_s(2~\mbox{GeV}) = 100-130$~MeV~\cite{QCDSF-UKQCD}, 
the ALPHA collaboration value is $97(22)$~MeV~\cite{ALPHA}, 
and the Rome value is
$101(8)(^{+25}_{-9}$)%{\scriptstyle +25 \atop \scriptstyle -9}
~MeV~\cite{Rome}.

The next major evolution in our lattice determination of the light quark masses will come from using a third lattice 
spacing, the ``super-fine'' configurations already partially 
implemented by MILC, and planned by UKQCD.  This will hopefully reduce 
the size of the chiral and systematic errors due to taste-breaking, and
vastly improve the quality of the continuum extrapolation.
The HPQCD collaboration plans to continue its accurate determinations 
of the fundamental constants by generalising this calculation in the first instance to heavy quark masses to complete the strong sector of the
Standard Model Lagrangian with two-loop 
calculations of all the quark masses with the already finished three-loop strong coupling constant.  
Further generalisation by insertion of appropriate operators will give 
the two-loop vector and axial-vector renormalisations, necessary for improving the accuracy of the decay constants $f_B$, $f_D$ etc and associated form-factors.  Eventually 
HPQCD plans therefore to have two-loop accurate determinations of many of the CKM matrix elements in the next few years.  

\acknowledgments
This work was supported by the US Department of Energy, the US National Science Foundation, the Natural Science and Engineering Research Council of Canada,
the Particle Physics and Astronomy Research Council of the UK.

%%%%%%%%%%%%%%%%%%%%%%%%%%%%%%%%%%%%
\appendix
\section{Perturbative Structure}
We present some details of the lattice size of the pole mass renormalisation
factor,~\eqref{e:LatticeMpole}. We define the full lattice quark 
propagator to be:
\newcommand{\pslash}{\widehat{p}\llap{$\slash$}}
\begin{align}
G(p,m_0)^{-1} = i\pslash + m_0 + \Sigma_{\rm tot}(p),
\label{e:fullprop}
\end{align}
where $\Sigma_{\rm tot}$ is minus the usual 1PI truncated two-point function. The 
lattice dispersion relation implied by $\hat p$ will be kept as a general 
function of the lattice momentum $p$; for unimproved staggered
quarks, for instance, $\hat p_\mu = \sin(p_\mu)$. 
Make the spinor decomposition:
\begin{align}
\Sigma_{\rm tot}(p) = \Sigma_1(p) + (i\pslash+m_0)\,\Sigma_2(p),
\label{e:sigma}
\end{align}
where both $\Sigma_1$ and $\Sigma_2$ are implicitly functions of $m_0$,
and both are Dirac scalars. 
Note that $\Sigma_2$ is only part of the wavefunction renormalisation.  
The pole mass is defined by the all-orders on-shell condition, corresponding
to the location of the pole in the propagator. At tree-level this is 
``$-i\pslash=m_0$'' 
(a very common but abusive notation). As is conventional, we work at zero external three momentum, $p=(p_t,\vec{0})$, and use the positive energy spinor projector (Euclidean), $(1+\gamma_t)/4$. Then it is convenient to 
rearrange~\eqref{e:fullprop} and~\eqref{e:sigma} to get
\begin{align}
P(p_t)\equiv-i\widehat{p}_t = m_0 + \frac{\Sigma_1(p)}{1+\Sigma_2(p)}
\label{e:mainsolve}
\end{align}
This relation can also be applied to theories with an additive mass renormalisation, such as Wilson quarks, by absorbing the momentum dependence of the additive mass into $P(p_t)$.
  
We recursively solve~\eqref{e:mainsolve} for the on-shell pole mass $M$,
defined by the renormalised energy at zero three-momentum
\begin{align}
 p_t=iM,\qquad\text{where}\;M=M^{(0)}+g_0^2 M^{(2)}+g_0^4 M^{(4)},
\label{e:ptM}
\end{align}
%with a perturbative expansion
%\begin{align}
% M=M^{(0)}+g_0^2 M^{(2)}+g_0^4 M^{(4)},
%\end{align}
is the perturbative expansion, with a corresponding series  for other quantities (including e.g.\
$p_t^{(0)} = i M^{(0)}$ and $P(iM^{(0)}) = m_0$).

We first consider the expansion of~\eqref{e:mainsolve} through one-loop, 
which reads
\begin{align}
P(iM^{(0)} + i g_0^2 M^{(2)}) = m_0 + g_0^2\Sigma_1^{(2)}(p^{(0)}) .
\end{align}
Hence
\begin{align}
M^{(2)}&=Z_\Psi^{(0)}\,\Sigma_1^{(2)}(p^{(0)}) ,\quad \text{where}\; {Z_\Psi^{(0)}}^{-1} =\left.\frac{d P(ix)}{dx}\right|_{x=M^{(0)}},
\label{e:M2}
\end{align}
is the tree-level wave function residue (e.g.\ 
$Z_\Psi^{(0)} = 1/\cosh(M^{(0)})$ for unimproved staggered quarks). 

To find the location of the pole at two-loops requires solving 
\eqref{e:fullprop} self-consistently
\begin{align}
P(iM^{(0)}+ i g_0^2 M^{(2)}+ig_0^4 M^{(4)}) &= m_0 + g_0^2 m^{(2)} + g_0^4 m^{(4)}\notag ,
\end{align}
where the right-hand side above represents a self-consistent expansion of 
the right-hand side of~\eqref{e:mainsolve}. Part of the $O(g_0^4)$ term 
arises from the one-loop piece of $\SIGMA_1$ when it is evaluated at the 
one-loop-corrected on-shell energy $p_t$, determined by \eqref{e:ptM} and \eqref{e:M2}.
By Taylor expansion:
\begin{align}
\Sigma_{1}^{(2)}(p_t) = 
\Sigma_{1}^{(2)}(iM^{(0)})
+ig_0^2 \, M^{(2)} \left.\frac{\partial \Sigma_1^{(2)}(p_t)}{\partial p_t}\right|_{p_t=iM^{(0)}} +\mathcal{O}(g_0^4).
\end{align}
The result for the two-loop contribution to the pole mass is therefore
\begin{align}
M^{(4)}&= Z_\psi^{(0)} \left(
m^{(4)}  - \frac12(M^{(2)})^2 
{\left.\frac{d^2 P(ix)}{dx^2}\right|_{x=M^{(0)}}} \right)
\label{e:M4}
\end{align}
(the second-term above is a correction of $O((am_0)^4)$ for Asqtad), and
\begin{align}
\label{e:littlem4}
m^{(4)} = \Sigma_1^{(4)}(iM^{(0)}) +
\Sigma_1^{(2)}(iM^{(0)}) \, Z_\psi^{(0)}  
\left\{\Tr\left[\frac{1+\gamma_t}4\frac{\partial\Sigma_{\rm tot}^{(2)}(p_t)}{\partial p_t}\right]_{p_t=iM^{(0)}}
\right\},
\end{align}
where the identity %The expression in curly brackets of~\eqref{e:littlem4} is what we used, but can be expanded in the continuum  to 
\begin{align}\label{e:continuumwavefn}
Z_\psi^{(0)}\Tr\left[\frac{1+\gamma_t}4\frac{\partial\Sigma_{\rm tot}^{(2)}(p_t)}{\partial p_t}\right]_{p_t=iM^{(0)}} =
-\Sigma_2^{(2)}(iM^{(0)}) + \left.i\frac{\partial\Sigma_1^{(2)}(p_t)}{\partial p_t}\right|_{p_t=iM^{(0)}}
\end{align}
was used.  In the continuum, $Z_\psi^{(0)}=1$ and~\eqref{e:continuumwavefn} is the complete one-loop expression for the wave function renormalisation there.
On the lattice, which has a tree level wavefunction renormalisation the result has some additional terms:
\begin{align}
Z_\Psi^{-1} = {Z_\Psi^{(0)}}^{-1} +g_0^2\Biggl\{&i\Tr\left[\frac{1+\gamma_t}4\frac{\partial\Sigma^{(2)}_{\text{tot}}(p)}{\partial p_t}\right]_{p=p^{(0)}}+M^{(2)}\left.\frac{d^2P(ix)}{dx^2}\right|_{p=p^{(0)}}\Biggr\}\notag.
\end{align}
Unfortunately~\eqref{e:continuumwavefn} is IR divergent for both lattice and continuum.  
This infrared divergence precisely cancels 
against an IR divergence in the two-loop nested-rainbow diagrams 
(D21, D22 and CT4 in figure~\ref{f:QMO2diags}). This parallels the cancellation of an 
infrared divergence in the continuum two-loop diagram, D21, 
\begin{fmffile}{QMO2_feyn}
\begin{fmfgraph}(75,30)
\fmfpen{thick}%
\fmfipair{l,r,li,ri,lii,rii,ci}%
\fmfiequ{r}{(0.95*w,0.05*h)}%
\fmfiequ{l}{(0.05*w,0.05*h)}%
\fmfiequ{li}{(0.2*w,0.05*h)}%
\fmfiequ{ri}{(0.8*w,0.05*h)}%
\fmfiequ{lii}{(0.33*w,0.05*h)}%
\fmfiequ{rii}{(0.66*w,0.05*h)}%
\fmfiequ{ci}{(0.5*w,0.05*h)}%
\fmfi{plain,f=blue}{l--li}%
\fmfi{plain,f=blue}{li--lii}%
\fmfi{fermion,f=blue}{lii--rii}%
\fmfi{plain,f=blue}{rii--ri}%
\fmfi{plain,f=blue}{ri--r}%
\fmfi{photon,f=green}{vbuild_arc(-1,li,ri)}%
\fmfi{photon,f=green}{vbuild_arc(-1,lii,rii)}%
\fmfiv{de.sh=circle,de.fil=full,de.si=2thick,f=black}{li}%
\fmfiv{de.sh=circle,de.fil=full,de.si=2thick,f=black}{ri}%
\fmfiv{de.sh=circle,de.fil=full,de.si=2thick,f=black}{lii}%
\fmfiv{de.sh=circle,de.fil=full,de.si=2thick,f=black}{rii}%
\end{fmfgraph},
which is likewise rendered finite by the iteration of the the one-loop 
self-energy, which generates a ``counterterm'' given by the one-loop mass 
shift times the one-loop wave function residue, analogous to the terms
in~\eqref{e:littlem4}.  

In this connection, we note that the continuum
pole mass was shown to be infrared finite at two-loops by Tarrach 
\cite{Tarrach:1980up}; an all-orders proof of the finiteness of the 
on-shell self-energy has only recently been established, by 
Kronfeld~\cite{Kronfeld:1998di}. 

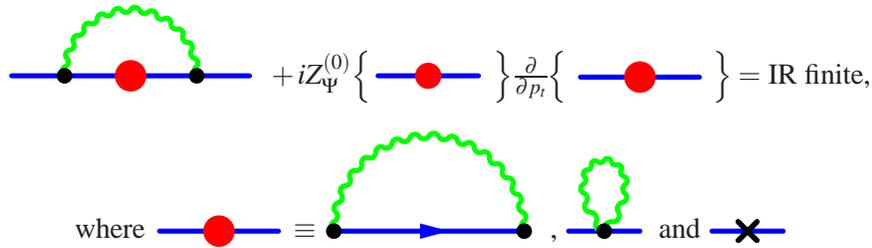
\begin{figure}
\begin{center}
\begin{fmfgraph}(100,60)
\fmfpen{thick}%
\fmfipair{l,r,li,ri,ci}%
\fmfset{wiggly_len}{2.5mm}%
\fmfiequ{r}{(0.95*w,0.05*h)}%
\fmfiequ{l}{(0.05*w,0.05*h)}%
\fmfiequ{li}{(0.25*w,0.05*h)}%
\fmfiequ{ri}{(0.75*w,0.05*h)}%
\fmfiequ{ci}{(0.5*w,0.05*h)}%
\fmfi{plain,f=blue}{l--li}%
\fmfi{fermion,f=blue}{li--ri}%
\fmfi{plain,f=blue}{ri--r}%
\fmfi{photon,f=green}{vbuild_arc(-1,li,ri)}%
\fmfiv{de.sh=circle,de.fil=full,de.si=2thick,f=black}{li}%
\fmfiv{de.sh=circle,de.fil=full,de.si=2thick,f=black}{ri}%
\fmfiv{de.sh=circle,de.fil=full,de.si=5thick,f=red}{ci}%
\end{fmfgraph}
$\!\!+\,iZ_\Psi^{(0)}\Bigl\{$
\begin{fmfgraph}(38,60)
\fmfpen{thick}%
\fmfipair{l,r,ci}%
\fmfiequ{r}{(w,0.05*h)}%
\fmfiequ{l}{(0,0.05*h)}%
\fmfiequ{ci}{(0.5*w,0.05*h)}%
\fmfi{plain,f=blue}{l--ci}%
\fmfi{plain,f=blue}{ci--r}%
\fmfiv{de.sh=circle,de.fil=full,de.si=4thick,f=red}{ci}%
\end{fmfgraph}
$\Bigr\}\frac{\partial}{\partial p_t}\!\Bigl\{$
\begin{fmfgraph}(50,50)
\fmfpen{thick}%
\fmfipair{l,r,ci}%
\fmfiequ{r}{(0.95*w,0.05*h)}%
\fmfiequ{l}{(0.05*w,0.05*h)}%
\fmfiequ{ci}{(0.5*w,0.05*h)}%
\fmfi{plain,f=blue}{l--ci}%
\fmfi{plain,f=blue}{ci--r}%
\fmfiv{de.sh=circle,de.fil=full,de.si=5thick,f=red}{ci}%
\end{fmfgraph}
\!$\!\Bigr\} = $ IR finite,\\
where
\begin{fmfgraph}(50,50)
\fmfpen{thick}%
\fmfipair{l,r,ci}%
\fmfiequ{r}{(0.95*w,0.05*h)}%
\fmfiequ{l}{(0.05*w,0.05*h)}%
\fmfiequ{ci}{(0.5*w,0.05*h)}%
\fmfi{plain,f=blue}{l--ci}%
\fmfi{plain,f=blue}{ci--r}%
\fmfiv{de.sh=circle,de.fil=full,de.si=5thick,f=red}{ci}%
\end{fmfgraph}
$\!\!\equiv$
\begin{fmfgraph}(80,50)
\fmfpen{thick}%
\fmfset{wiggly_len}{2.5mm}%
\fmfipair{l,r}%
\fmfiequ{r}{(0.95*w,0.05*h)}%
\fmfiequ{l}{(0.05*w,0.05*h)}%
\fmfi{fermion,f=blue}{l--r}%
\fmfi{photon,f=green}{vbuild_arc(-1,l,r)}%
\fmfiv{de.sh=circle,de.fil=full,de.si=2thick,f=black}{l}%
\fmfiv{de.sh=circle,de.fil=full,de.si=2thick,f=black}{r}%
\end{fmfgraph}
,
\begin{fmfgraph}(30,50)
\fmfpen{thick}%
\fmfipair{l,r,ci}%
\fmfset{wiggly_len}{2.5mm}%
\fmfipath{p}
\fmfiequ{r}{(0.95*w,0.05*h)}%
\fmfiequ{l}{(0.05*w,0.05*h)}%
\fmfiequ{ci}{(0.5*w,0.05*h)}%
\fmfi{plain,f=blue}{l--ci}%
\fmfi{plain,f=blue}{ci--r}%
\fmfiequ{p}{(ci){dir(45)}..(ci+2/3*40*dir(90))..{dir(-45)}(ci)}%
\fmfi{photon,f=green}{p}
\fmfiv{de.sh=circle,de.fil=full,de.si=2thick,f=black}{ci}%
\end{fmfgraph}
and
\begin{fmfgraph}(30,50)
\fmfpen{thick}%
\fmfipair{l,r,ci}%
\fmfiequ{r}{(0.95*w,0.05*h)}%
\fmfiequ{l}{(0.05*w,0.05*h)}%
\fmfiequ{ci}{(0.5*w,0.05*h)}%
\fmfi{plain,f=blue}{l--ci}%
\fmfi{plain,f=blue}{ci--r}%
\fmfiv{de.sh=cross,de.fil=full,de.si=5thick,f=black}{ci}%
\end{fmfgraph}
\caption{\label{f:IRsub}Schematic representation of an IR subtraction,
where appropriate traces of the self-energy with an energy-projector 
are implicit in this schematic representation.}
\end{center}
\end{figure}

On the lattice, the infrared cancellation in~\eqref{e:littlem4} has
a schematic representation in diagrammatic form, as shown in
figure~\ref{f:IRsub}. 
We numerically evaluate the integral for the two-loop diagram on the left
in figure~\ref{f:IRsub} with a subtraction, in its integrand, of the 
product of the independent integrands for the one-loop 
self-energy, and its derivative. This grouping, as indicated in
figure~\ref{f:IRsub}, is IR finite, and does not require any infrared regulator. 
We obtain a powerful check of this result by noting that this combination
generates a leading logarithmic contribution to the anomalous dimension 
of the mass which goes like $\log^2(a^2m_0^2)$, and whose coefficient is 
identical to that of the infrared-subtracted combination in the continuum,
which can be found in Feynman gauge from the \msbar\ results 
in~\cite{Tarrach:1980up}.  

All the diagrams for the two-loop self-energy, figure~\ref{f:QMO2diags}, were
generated and evaluated independently by two of us. The Feynman rules for
the highly-improved actions are exceedingly complicated, and were
generated automatically using computer algebra-based codes 
\cite{Trottier:2003bw}.
Moreover, one of us has produced an algorithm which automatically generates
the Feynman diagrams themselves~\cite{Mason:2004zt}. The two-loop integrals
were evaluated numerically, using the adaptive Monte-Carlo sampling of 
method of {\tt VEGAS}~\cite{vegas}.  Practically, when performing the subtractions of figure~\ref{f:IRsub} the integrals are much more convergent with a (4D) spherical transform and a further $k_{\text{new}}=\log|k|$ transformation.  The effect of the associated Jacobian's is to regulate the IR with $k^4$ and a spherical cut-off at small $|k|$.  We found that $e^{-10}$ and smaller were sufficient to show the integrals independent of this cut-off.  Almost all diagrams benefited from using the spherical transform, though some of the other ``continuum'' like diagrams are better behaved with the ``log-spherical'' transform for very large numbers of integrand evaluations.  We have checked that varying the cut-off makes no difference to the result.    
A powerful additional cross-check was provided by an explicit verification 
that our results are gauge independent, which we established numerically for 
two different bare-quark masses in three covariant gauges: Feynman, Landau 
and Yennie.

\begin{figure}[t]
\begin{center}
\psfrag{ylabel}[cc][][1.5]{\large $A_{20}|_{n_f=0}$ in $\alpha_L^2$}
\psfrag{TOT Imp}[cc][][1.5]{\large Asqtad multiplicative mass renormalisation at two-loops}
\psfrag{Quark Mass}[cc][][1.5]{\large Quark Mass $(m_0 a)$}
\includegraphics[angle=-90,width=5in,clip]{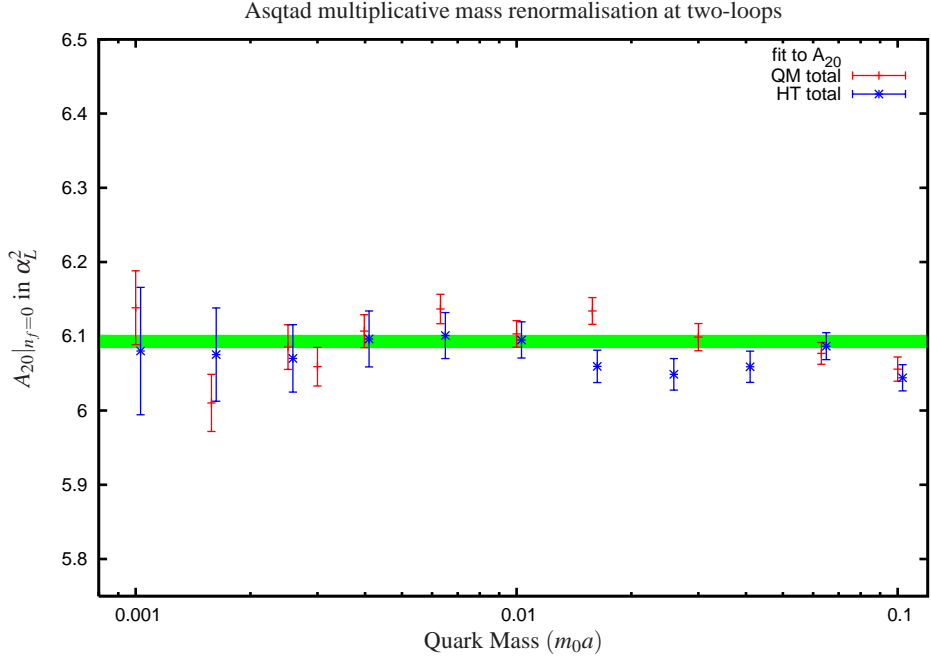}
\caption{\label{f:results}The total of the gluonic part of the two-loop contribution to the pole mass on the lattice with the Asqtad action, evaluated independently by two of the authors, in units of $\alpha_L^2$, for varying quark mass 
$(m_0 a)$. The data is shown \emph{after} subtracting the known logarithms: $A_{22} \log^2 m_0a + A_{21} \log m_0a$, and should therefore be the constant, $A_{20}$, with no mass dependence.  The absence of observed lattice artifacts of the form $(m_0 a)^n\log^l m_0 a$, $l\leq2$ justifies our assumption in the derivation that for small masses the lattice is a mass-independent renormalisation scheme.}
\end{center}
\end{figure}

We show results for the two-loop part of the pole mass,~\eqref{e:LatticeMpole}, coming from diagrams without fermion loops, 
in figure~\ref{f:results}. As described in Sect. 2, we can test our
calculation by subtracting the known logarithms in $m_0a$, in order to 
expose the remaining factor $A_{20}$, which must be finite in the
limit $m_0 a \to 0$. Figure~\ref{f:results} shows results for the
gluonic part of $A_{20}$ over a wide range of bare masses, which
clearly shows the expected limiting behaviour. 

\begin{figure}
\begin{center}
\includegraphics[angle=90,width=4.5in]{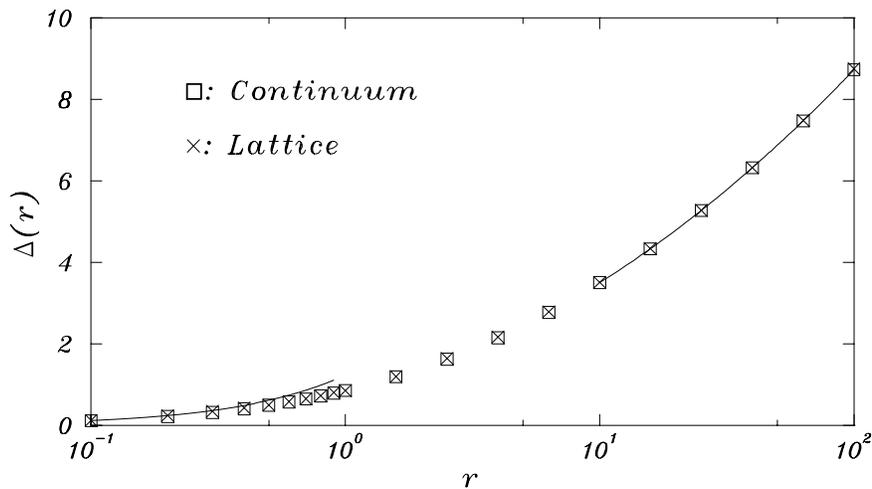}
\caption{\label{f:deltar}Comparison of the sea loop-quark mass
dependence of pole mass renormalisation matching factors, on the 
lattice-side of the matching, and on the continuum side. The difference between the continuum squares and lattice crosses (calculated at $a m_{\text{valence}}=0.001$), is smaller than the errors.  
The variable
$r=m_{\rm sea} / m_{\rm valence}$. The solid lines show limiting
forms of the dilogarithmic but analytic continuum function.%, cf.\ \ (\protect\ref{e:Deltall})--(\protect\ref{e:Deltagg}).
}
\end{center}
\end{figure}

A further stringent check of our evaluation of the diagrams with
internal fermion loops (diagrams 12, 13, 19 and 20 in 
figure~\ref{f:QMO2diags}) is achieved by computing over a range of 
sea quark masses. As described in Sect.~3, the mass dependence in the
intermediate renormalisation from the bare mass to the pole mass $M$
should cancel against the renormalisation from $M$ to the \msbar\ mass.
We define
\begin{align}
  A_{20}(r_{\rm sea}) \equiv A_{20}(0) 
+ \frac4{3\pi^2} \Delta_{\rm lattice}(r_{\rm sea}) ,
  \quad r_{\rm sea} = \frac{m_{\rm sea}}{m_{\rm valence}} ,
\end{align}
and compare with the analogous continuum function $\Delta(r_{\rm sea})$.
%cf.\ Eqs. (\ref{e:msbarMpole})--(\ref{e:Deltagg})
 We plot our results in 
figure \ref{f:deltar}, over a very wide range in $r_{\rm sea}$,
which shows the expected cancellation.

\end{fmffile}

\end{document}